\documentclass[letterpaper,aps,prl,twocolumn,superscriptaddress,showpacs]{revtex4}
\usepackage{graphicx}
\usepackage{amsmath}
\usepackage{amssymb}

\begin{document}
\title{Nonlinear modal interactions in clamped-clamped mechanical resonators}
\author{H.J.R. Westra}\email{h.j.r.westra@tudelft.nl}
\author{M. Poot}
\author{H.S.J. van der Zant}
\author{W.J. Venstra}
\affiliation{Kavli Institute of Nanoscience, Delft University of Technology, Lorentzweg 1, 2628CJ Delft, The Netherlands}
\date{\today}
\begin{abstract}
A theoretical and experimental investigation is presented on the intermodal coupling between the flexural vibration modes of a single clamped-clamped beam. Nonlinear coupling allows an arbitrary flexural mode to be used as a self-detector for the amplitude of another mode, presenting a method to measure the energy stored in a specific resonance mode. Experimentally observed complex nonlinear dynamics of the coupled modes are quantitatively captured by a model which couples the modes via the beam extension; the same mechanism is responsible for the well-known Duffing nonlinearity in clamped-clamped beams.
\end{abstract}
\maketitle
\begin{figure}[!h!t]
\includegraphics[width=80mm]{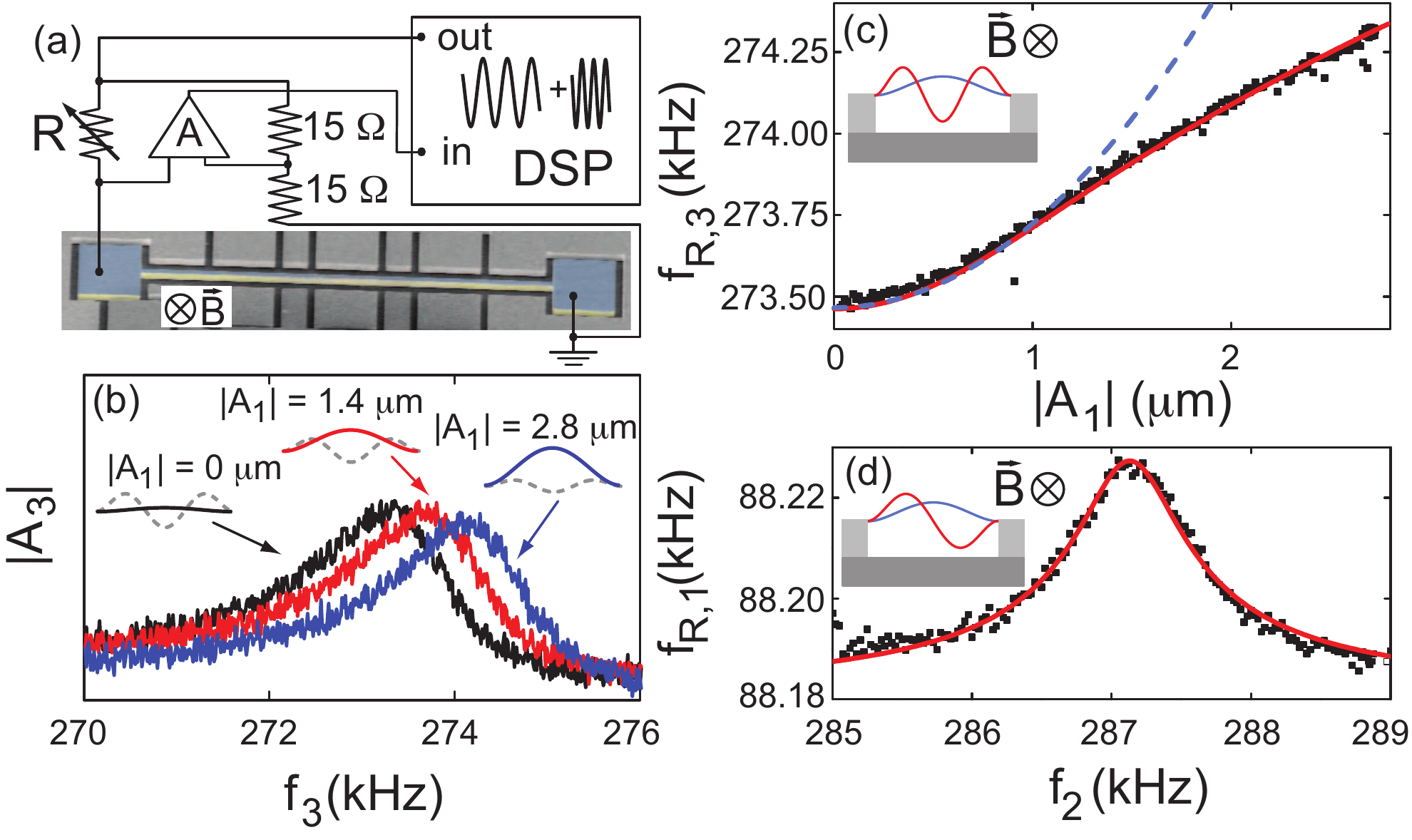}
\caption{(a) Setup, with a colored scanning electron micrograph of the resonator beam. DSP is the digital signal processor, R is used to balance the bridge and set to 22 $\Omega$ and A is the amplifier. (b) Frequency responses of the third mode (amplitude $|A_3|$) for different drive amplitudes of the first mode $|A_1|$ on resonance. The drive current of mode 3 is $I_3 = 1.5$ mA. Inset are the beam shapes of the first (solid) and third mode (dashed) for increasing amplitude of the first mode. For clarity, the amplitude of the third mode has been lowered more than in the actual situation. (c) Resonance frequency of mode 3 ($f_{\mathrm{R},3}$) as a function of drive amplitude of mode 1 with measured data (black squares) and result from the model (red line). For small drive currents a quadratic dependence is observed (dashed blue line). Inset: schematic of the resonator in a magnetic field showing the beam shape of the first (blue) and third (red) mode. (d) On a different device, the resonance frequency of the first mode is used as a detector of the second mode. Inset: beam shape of the first (blue) and second (red) mode. The error bars are  within in data markers.}
\label{fig:1}
\end{figure}
\begin{figure}[!h!t]
\includegraphics[width=85mm]{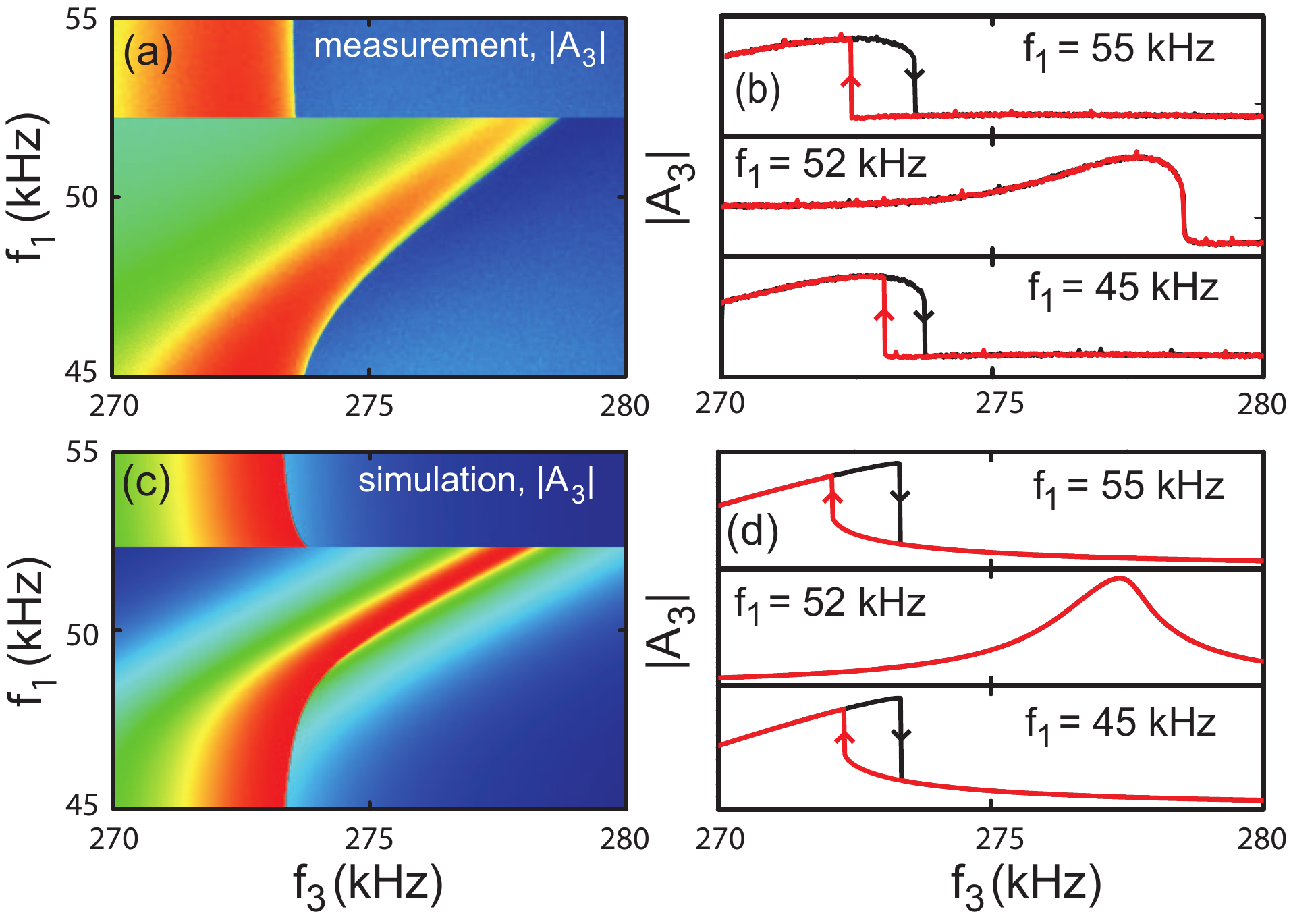}
\caption{(a) Frequency-frequency plot for the forward sweep of the third mode. Red indicates a high amplitude and blue corresponds to a low amplitude. On resonance of the first mode, the third mode is linear and off-resonance it is nonlinear, which demonstrates the tuning of the nonlinearity. (b) Forward (black lines) and back (red lines) sweeps are shown below resonance, on resonance and above resonance, with $f_1 = 45, 52$ and 55 kHz respectively. Driving currents are $I_1 = 2$ mA  and $I_3 = 8$ mA. (c) Result of the simulation with parameters extracted from the experiment. (d) Simulated traces for the situation in (b).}
\label{fig:3}
\end{figure}

An important topic in nanomechanics is the motion detection of mechanical resonators. Several schemes have been proposed to attain sensitivities near the quantum limit of mechanical motion~\cite{Schwab:2005p36}, whereas application-driven research is focussed on on-chip detection~\cite{Li:2007p114} and readout of resonator arrays~\cite{Venstra:2008p379}. Central in any detection scheme is the coupling of a mechanical resonator to another system, which transduces the motion into a measurable quantity. Examples of sensitive detectors include a single-electron transistor~\cite{LaHaye:2004p1464}, a microwave cavity~\cite{Regal:2008p555}, or an optical interferometer~\cite{Kippenberg:2008p7}. A second mechanical resonator can also be used to detect the motion of the resonator~\cite{Buks2002,Karabalin:2009p177}. Such a system of coupled resonators has been proposed as a quantum nondemolition detection scheme, in which one resonator is in a quantum state~\cite{Santamore:2004p398}. Coupling between different mechanical resonators is often present in large-scale integrated arrays due to electrostatic~\cite{Buks2002} and mechanical interaction~\cite{Karabalin:2009p177}. Coupling between individual resonators can also lead to complex behavior, which is theoretically well-documented~\cite{Lifshitz2003}.\\
\indent \indent In this Letter, we study the coupling between vibrational modes in a \emph{single} beam resonator. We demonstrate that flexural modes are coupled by the displacement-induced tension in the beam. Using this coupling, the displacement of any mode can be detected by measuring the response of another mode, making otherwise undetectable modes visible. We present a general theoretical framework based on the Euler-Bernoulli equation extended with displacement-induced tension. The model quantitatively describes the complex dynamic behavior observed in the regime where two modes are simultaneously driven nonlinear. The coupling mechanism plays an prominent role in the dynamics of carbon nanotube resonators and resonators under high tension, and should be taken into account when describing such systems accurately. \\
\indent \indent Experiments are performed on a single-crystalline silicon beam with dimensions $L \times w \times h = 1000 \times 35 \times 6 \, \mu\mathrm{m}^3$ fabricated by patterning a silicon-on-insulator wafer and subsequent wet etching. The resonator is placed in a magnetic field of $B$ = 2.1 T and a magnetomotive technique~\cite{Venstra:2008p379,Yurke:1995p872} is used to detect the mechanical motion of the beam at room temperature and atmospheric pressure (see Figure~\ref{fig:1}a). The beam is driven at multiple frequencies by sending alternating currents through a conductive aluminum path, evaporated on top of the resonator. The motion of the beam in the magnetic field generates an electromotive force, which is balanced using a Wheatstone bridge, amplified, and then digitized (see Figure~\ref{fig:1}a). The frequency response (amplitude and phase) of the resonator at the two drive frequencies is calculated using a digital signal processor.\\
\indent \indent Measurements are conducted on the first and third flexural mode of the beam. In Fig.~\ref{fig:1}b the frequency response of the third mode is shown for three different drive amplitudes of the first mode. The resonance frequency of the third mode increases when the amplitude of the first mode $|A_1|$ becomes larger. At the same time the amplitude decreases slightly. The resonance frequency $f_{\mathrm{R},3}$, obtained by fitting a damped driven harmonic oscillator response, is plotted as a function of the drive amplitude of the first mode in Fig.~\ref{fig:1}c. For small amplitudes $|A_1|$, a quadratic dependence of $f_{\mathrm{R},3}$ on $|A_1|$ is found (dashed line in Fig.~\ref{fig:1}c). To give a qualitative picture, which explains the observed coupling in Figure~\ref{fig:1}b. When driving the third mode on resonance, the tension in the beam increases as the amplitude of the first mode increases. This results in a higher frequency and a lower amplitude of the third mode. \\
\indent \indent As mentioned above, using the magnetomotive measurement scheme it is not possible to detect the even resonance modes. However, by exploiting the coupling to detectable odd modes, their motion is observable. We have detected the motion of the second mode by measuring its influence on the first mode. To drive the second mode, the sample (different device, $h = 10 \, \mu$m) is mounted on a piezo actuator and excited at frequencies around the second resonance mode. Off-resonance, no frequency shift of the first mode is observed. When driven at the resonance frequency of the second mode, the resonance frequency of the first mode shifts to a higher value (Fig.~\ref{fig:1}d). The response of the second mode is obtained by measuring the shift in $f_{\mathrm{R},1}$, whereas a magnetomotive measurement around the same frequency shows no signature of the second mode. For weak piezo driving the frequency shift is proportional to $|A_2|^2$ and a squared damped driven harmonic oscillator function fits the data well with $f_2 = 287$ kHz and $Q = 250$.\\
\indent \indent In order to quantify the coupling between the flexural modes of the beam, an analytical model is developed. First, the equations are derived for the general situation with modes coupled; then we focus on the experimental situation, where only two modes are considered. The Euler-Bernoulli equation including tension $\mathcal{T}$~\cite{Sapmaz:2003p389,Nayfeh:1995p387,Poot:2007p392} is used as starting point. To simplify the notation, the displacement $u$ is scaled with the beam thickness $h$, and the coordinate $x$ with the beam length $L$. A displacement of the beam causes an elongation and increases the tension. The dimensionless tension $T = L^2\mathcal{T}/D$ ($D$ is the bending rigidity) is given by:
\begin{equation}
T = T_{0} + \frac{\tau}{2}\int_0^1 \Big(\frac{\partial u}{\partial x}\Big)^2 \mathrm{d}x.
\label{eq:tension} 
\end{equation}
$T_0$ is the residual tension in the beam and $\tau = h^2A/I_y$, with $I_y$ the second moment of inertia and $A$ the cross-section. For a rectangular beam $\tau$ equals 12~\cite{NCleland:2003p509}. The displacement $u$ and dimensionless force $F = L^4\mathcal{F}/Dh$ can be split into a dc and an ac part i.e., $u = u_{\mathrm{dc}} + u_{\mathrm{ac}}$ and $F = F_{\mathrm{dc}} + F_{\mathrm{ac}}$. This yields a well-known equation for the static displacement~\cite{Sapmaz:2003p389,Nayfeh:1995p387,Poot:2007p392} and an equation for the ac motion:
\begin{equation}
\ddot{u}_{\mathrm{ac}} + \eta \dot{u}_{\mathrm{ac}} + \mathcal{L}[u_\mathrm{ac}] - (T - T_{\mathrm{dc}})u_{\mathrm{ac}}'' - (T - \overline T  -T_{\mathrm{ac}})u_{\mathrm{dc}}'' = F_{\mathrm{ac}}.
\label{eq:actension}
\end{equation}  
Here, $T_{\mathrm{dc}}$ is the residual tension plus the tension from the dc displacement and $\overline T$ is the time-averaged tension, which also contains terms proportional to $u_{\mathrm{ac}}^2$. For small ac displacements $\overline T \approx T_{\mathrm{dc}} $. $T_{\mathrm{ac}}$ contains all terms that are linear in $u_{\mathrm{ac}}$. The operator $\mathcal{L}[u]$ is defined as~\cite{Poot:2007p392}:
\begin{equation}
\mathcal{L}[u] = u'''' - T_{\mathrm{dc}}u'' - T_{\mathrm{ac}}[u]u_{\mathrm{dc}}''.
\label{eq:L}
\end{equation}
The first three terms on the left side of Eq.~\ref{eq:actension} determine the linear response of the system. The nonlinearity is introduced with $u_{\mathrm{ac}}^2$ and $u_{\mathrm{ac}}^3$, which occur in the last two terms. \\
\indent \indent The resonance frequency for infinitesimal small amplitudes $\omega_{0,i}^2$ and the corresponding shape $\xi_i(x)$ of mode $i$ are the eigenvalues and orthonormal eigenfunctions of $\mathcal{L}$ respectively. The ac displacement is expanded in terms of the mode shapes as $u_\mathrm{ac} = \sum_{i=1}^{\infty} u_i(t)\xi_i(x)$. The dc displacement is defined as $u_{\mathrm{dc}}  \equiv u_{0}\xi_0(x)$ and $\int_0^1 \xi_i'(x)\xi_j'(x)\mathrm{d}x$ is denoted as $I_{ij}$. The value of the integral $I_{ij}$ depends only on the shapes of mode $i$ and $j$ and and can be calculated numerically.
Using this notation the tension $T$ is:
\begin{equation}
 T(t)= T_{0} + \frac{\tau}{2}\sum_{i,j=0}^{\infty} u_{i}(t)u_{j}(t)I_{ij},
\label{eq:tension2}
\end{equation}
so that $T_{\mathrm{dc}} = T_{0} + \frac{\tau}{2} u_0^2 I_{00}$ and $T_{\mathrm{ac}} = \frac{1}{2}\tau u_0 \sum_{n=1}^\infty u_{n}I_{n0}$. For nonzero amplitudes $u_n$ the tension increases, tuning the resonance frequencies $\omega_{\mathrm{R},i}$ away from $\omega_{0,i}$. The effect of the resonance frequency on its own motion results in a Duffing equation (i.e., the resonance frequency increases with its own amplitude). Based on the same concept, the tuning of the resonance frequency due to the motion of other modes can be envisaged and this coupling is the central theme of this work. \\
\indent \indent The displacement of a beam, which is driven at frequencies $\omega_i$ is written as:
\begin{equation}
u(x,t) = \sum_i |a_i| \xi_i(x) \mathrm{cos}(\omega_i t + \angle a_i),
\label{eq:uac}
\end{equation}
where the $a_i = A_i / h$ are the complex amplitudes of the mode at $\omega_i$. Substituting Eq.~\ref{eq:uac} into Eq.~\ref{eq:tension2} gives an expression for the total tension. The two tension terms in the ac equation are given by:
\begin{widetext}
\begin{eqnarray}
T-T_{\mathrm{dc}} &=& \frac{\tau }{2}\sum_{i>0} \Big( \frac{1}{2}|a_i|^2 I_{ii}+ \frac{u_0}{2} a_i e^{i\omega_i t} I_{0i} + \mathrm{c.c.}\Big)+ (T- \overline{T}- T_{\mathrm{ac}}) \nonumber \\
T- \overline{T}- T_{\mathrm{ac}} &=& \frac{\tau}{4}\sum_{i\neq j > 0} \Big(\frac{1}{2} a_i^2 e^{i2\omega_i t} I_{ii} +  a_i a_j^* e^{i(\omega_i - \omega_j)t}I_{ij} + a_i a_j e^{i(\omega_i + \omega_j)t}I_{ij}+ \mathrm{c.c.} \Big).
\label{eq:tensionterms}
\end{eqnarray}
\end{widetext}
Here, c.c. stands for the complex conjugate of the term before c.c. between the round brackets. The expressions for the tension, Eq.~\ref{eq:tensionterms}, are substituted in the equation of motion, Eq.~\ref{eq:actension}. The time-averaged equation for the amplitude of mode $i$ driven at frequency $\omega_i$, is then given by:
\begin{widetext}
\begin{eqnarray}
\sum_{i\neq j>0}\Big\{ \Big( \omega_{0,i}^2- \omega_i^2 + i\omega_i\omega_{0,i}/Q_i+ \frac{\tau}{4}|a_i|^2 I_{ii}^2 + \frac{\tau}{4}\Big(|a_j|^2I_{ii}I_{jj} +|a_j|^2I_{ij}^2 \Big)\Big)a_i -  \int_0^1 F_{\mathrm{ac}} \xi_i \mathrm{d}x\Big\} = 0,
\label{eq:solution} 
\end{eqnarray}
\end{widetext}
where $F_{\mathrm{ac}} = L^4 \mathcal{F}_{\mathrm{ac}}/Dh$ is the dimensionless ac force.
\indent \indent So far, the analysis is valid for any flexural resonator. We now focus on the experimental situation where, unlike in buckled beams~\cite{Nayfeh:1995p387} and string-like resonators~\cite{Verbridge:2008p013112}, the residual tension does not play a significant role, $T_0 = 0$. Moreover, the static displacement $u_0 = 0$, which may not be the case for carbon nanotube resonators~\cite{Sazonova:2004p4487}, where a gate voltage induces a static displacement. We finally assume that the resonances are resolvable: $|\omega_{\mathrm{R},i} - \omega_{\mathrm{R},j}| \gg \omega_{\mathrm{R},i}/Q_i + \omega_{\mathrm{R},j}/Q_j$ for $i \neq j$. The homogeneous force per unit length on the beam is $\mathcal{F}_{\mathrm{ac}}= B I$, where $I$ is the current through the resonator. \\
\indent \indent To compare the model with the data in Fig.~\ref{fig:1}c, we first extract the experimental values of the parameters. The resonance frequencies for the first and third mode are $f_1 = 48.2$ kHz and $f_3 = 273.4$ kHz, close to the predicted values of 46.1 and 249 kHz for a beam-like resonator. Their Q-factors are $Q_1 = 41 $ and $Q_3 = 172 $. The values of $I_{ij}$ determine the coupling strength: $I_{11} = 12.3$, $I_{33} = 98.9$ and $I_{13} = I_{31} = -9.7$. The average displacement of the modes per unit deflection, $\int_0^1 \xi_i(x) dx$, are $0.83$ and $0.36$ for mode 1 and 3 respectively. The model is solved numerically by calculating the amplitudes of the two modes self-consistently for the experimental conditions and without any free parameters. The calculated resonance frequency of the third mode as a function of the amplitude of the first mode is shown in Fig.~\ref{fig:1}c. Excellent agreement is found between the observed frequency shift and the prediction by the model. The proposed model which couples modes though tension induced by the vibrations is therefore a good description of the experiment. For large amplitudes, the resonance frequency scales with $|A_1|^{2/3}$, indicating that the beam is in the strong bending regime~\cite{Sapmaz:2003p389}. For small $|A_1|$, the tuning is quadratic and the third mode can be used to detect the amplitude of the first mode with a sensitivity of 0.18 Hz nm$^{-2}$, which is determined from the quadratic curve in Fig.~\ref{fig:1}c.\\
\begin{figure}[!h!t]
\includegraphics[width=80mm]{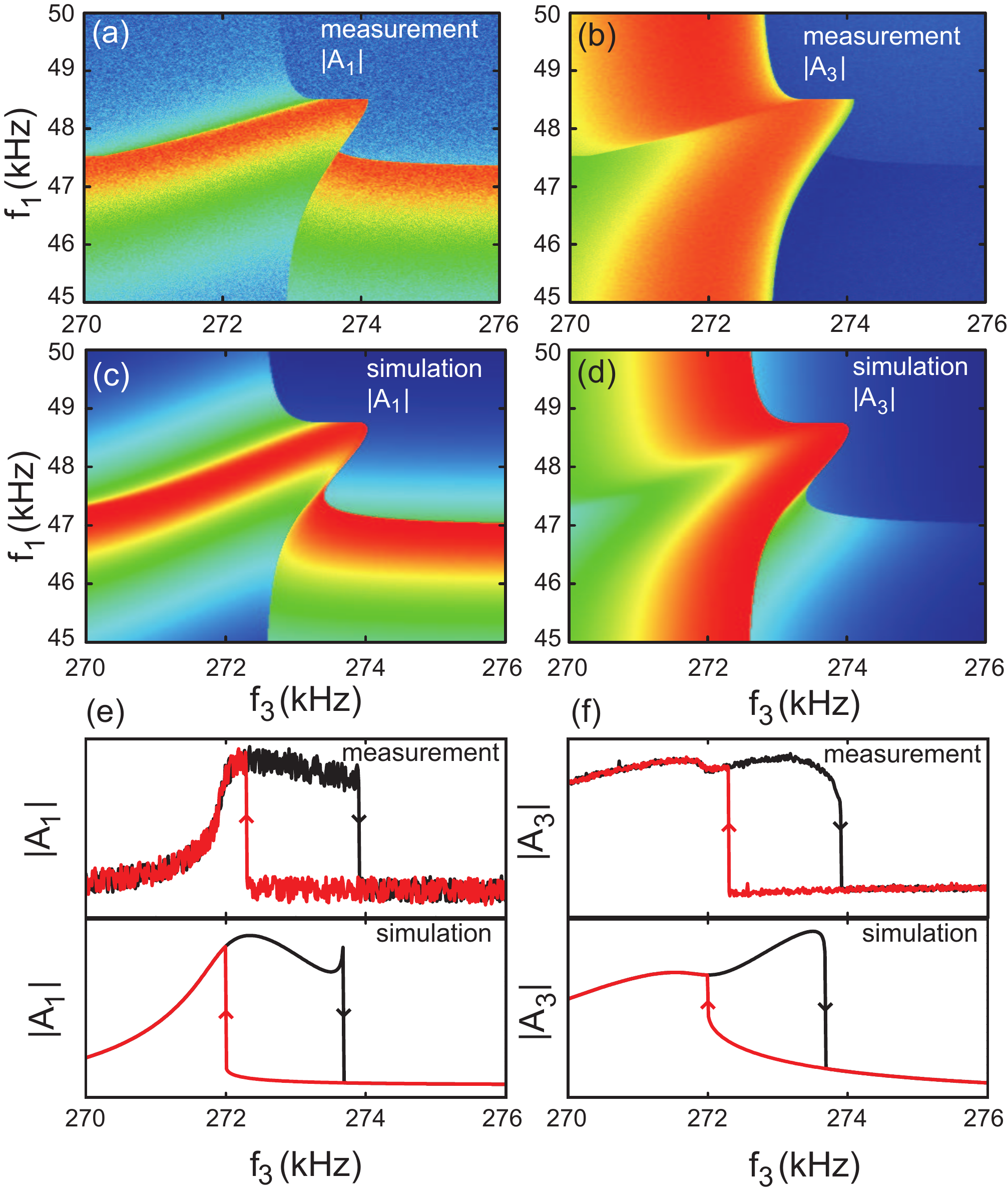}
\caption{Frequency-frequency response of the simultaneously driven and detected first and third mode. The drive frequency of the third mode is swept, while driving the first mode at a fixed frequency. After each sweep the frequency of the first mode is increased. The amplitudes $|A_1|$ (a) and $|A_3|$ (b) are recorded. The driving currents are $I_1 = 0.8$ mA  and $I_3 = 7.0$ mA. Red indicates a high amplitude and blue corresponds to a low-amplitude response. Simulations are shown in (c) and (d). The frequency response is plotted for the amplitude of mode 1 (e) and mode 3 (f) at $f_1 = 48$ kHz, also the back sweep is shown for the two cases.}
\label{fig:2}
\end{figure}    
\indent \indent To further test the consistency of our model, we study the complex dynamics of the coupled modes. When driving both modes nonlinear, interesting features are observed. In Fig.~\ref{fig:2}a the amplitude of the nonlinear first mode is plotted versus the driving frequencies $f_1$ and $f_3$. Simultaneously the amplitude of the third mode is recorded (Fig.~\ref{fig:2}b). The two modes interact with each other as the nonlinear line shape of one mode is reflected in the response of the other mode. Also a frequency response with two peaks, which is clearly different from a Duffing line shape, is observed as illustrated more clearly in Fig.~\ref{fig:2}e and f. The two peaks arise from the bistable first mode, where two values for the amplitude are possible. These two amplitudes correspond to two values of the tension, which leads to two resonance frequencies of the third mode and to two peaks in its frequency response. The simulation with the parameters as stated above, reproduces all observed features in the amplitude of both modes (Fig.~\ref{fig:3}c-d). This indicates that the model captures the coupling mechanism in detail~\cite{footnote}.  \\
\indent \indent An example of how the coupling between the modes can be used in practice, is the increase in dynamic range of the mechanical resonator. In small-scale resonators the dynamic range is limited by the nonlinear response at strong driving amplitudes, which is disadvantageous for many applications~\cite{Kraus:2001,Postma:2005p1338}. Our analysis and experiments show that there is a way to extend the dynamic range of one mode by driving another mode on resonance at high amplitudes. Fig.~\ref{fig:3}a,c shows the frequency response of the third mode when the frequency of the first mode is swept across its nonlinear resonance. Away from $f_{\mathrm{R},1}$ the third mode shows a Duffing-like response as illustrated in the top and bottom panel of Fig.~\ref{fig:3}b,d. However, when driving the first mode on resonance (middle panel) the third mode displays a hysteresis-free response. Effectively, the nonlinearity constant in the Duffing equation is decreased, which can be understood as follows: when the third mode enters its resonance, its amplitude increases and the increased tension tunes the resonance frequency of the first mode up. The amplitude of the first mode then drops, reducing the tension and lowering the resonance frequency of the third mode. This feedback mechanism reduces the cubic stiffness of the third mode and makes the third mode linear, thereby increasing the dynamic range. \\
\indent \indent The presented model applies to any clamped-clamped geometry, ranging from suspended bridges to carbon nanotubes. For nanomechanical devices with high aspect ratios, the flexural rigidity can be neglected and the restoring force originates from the axial rigidity. The energy of the resonator is stored in the beam elongation which couples directly to the ac tension. This increases the coupling between the resonance modes and makes the detection mechanism well-suited for nanowires and nanotubes. Furthermore in nanomechanical devices, significant residual tension ($T_0$) may be present in the beam. Only the numerical values $I_{ij}$ change in that case. For a resonator with high $T_0$, the off-diagonal elements $I_{ij} = 0$ and for mode 1 and 3, $I_{11}$ is 9.9 and $I_{33}$ is 88.8. Thus, for a string-like resonator the coupling term in Equation~\ref{eq:solution} remains within the same order of magnitude as in the beam resonator used in the experiment. To quantify the effect of modal interactions in a nanomechanical resonator with tension, we consider the suspended carbon nanotube with a high-quality factor from Ref.~\cite{huttel:2009}. Taken the parameters listed in that paper, we calculate the sensitivity of the third mode to the amplitude of the first mode. We find a value of 1 MHz nm$^{-2}$, which is more than six orders of magnitude larger than the value found from the dashed line in Fig.~\ref{fig:1}c.\\
\indent \indent In conclusion, we have measured the coupling between flexural modes of a clamped-clamped beam resonator by simultaneously driving the beam at multiple frequencies. We observe nonlinear interaction between the modes in the linear regime, and complex dynamics at large driving amplitudes. When describing the motion of a mechanical resonator, it is necessary to include this interaction, since this mechanism divides the available energy over the modes, and plays a role in the energy dissipation in the resonator. A theoretical model is developed, which couples arbitrary flexural modes via the tension. The model is in excellent agreement with the measurements and quantitatively captures the observed complex dynamics. The nonlinear coupling can be used to detect resonance modes that would otherwise be inaccessible by the experiment, to tune the nonlinearity constant, and to increase the dynamic range of micro- and nanomechanical resonators.  \\
\indent\indent The authors acknowledge financial support from the Dutch funding organizations FOM and NWO (VICI).


\begin{thebibliography}{18}
\expandafter\ifx\csname natexlab\endcsname\relax\def\natexlab#1{#1}\fi
\expandafter\ifx\csname bibnamefont\endcsname\relax
  \def\bibnamefont#1{#1}\fi
\expandafter\ifx\csname bibfnamefont\endcsname\relax
  \def\bibfnamefont#1{#1}\fi
\expandafter\ifx\csname citenamefont\endcsname\relax
  \def\citenamefont#1{#1}\fi
\expandafter\ifx\csname url\endcsname\relax
  \def\url#1{\texttt{#1}}\fi
\expandafter\ifx\csname urlprefix\endcsname\relax\def\urlprefix{URL }\fi
\providecommand{\bibinfo}[2]{#2}
\providecommand{\eprint}[2][]{\url{#2}}

\bibitem[{\citenamefont{Schwab and Roukes}(2005)}]{Schwab:2005p36}
\bibinfo{author}{\bibfnamefont{K.~C.} \bibnamefont{Schwab}} \bibnamefont{and}
  \bibinfo{author}{\bibfnamefont{M.~L.} \bibnamefont{Roukes}},
  \bibinfo{journal}{Physics Today} \textbf{\bibinfo{volume}{58}},
  \bibinfo{pages}{36} (\bibinfo{year}{2005}).

\bibitem[{\citenamefont{Li et~al.}(2007)\citenamefont{Li, Tang, and
  Roukes}}]{Li:2007p114}
\bibinfo{author}{\bibfnamefont{M.}~\bibnamefont{Li}},
  \bibinfo{author}{\bibfnamefont{H.~X.} \bibnamefont{Tang}}, \bibnamefont{and}
  \bibinfo{author}{\bibfnamefont{M.~L.} \bibnamefont{Roukes}},
  \bibinfo{journal}{Nature Nanotech.} \textbf{\bibinfo{volume}{2}},
  \bibinfo{pages}{114} (\bibinfo{year}{2007}).

\bibitem[{\citenamefont{Venstra and van~der Zant}(2008)}]{Venstra:2008p379}
\bibinfo{author}{\bibfnamefont{W.~J.} \bibnamefont{Venstra}} \bibnamefont{and}
  \bibinfo{author}{\bibfnamefont{H.~S.~J.} \bibnamefont{van~der Zant}},
  \bibinfo{journal}{Appl. Phys. Lett.} \textbf{\bibinfo{volume}{93}},
  \bibinfo{pages}{234106} (\bibinfo{year}{2008}).

\bibitem[{\citenamefont{LaHaye et~al.}(2004)\citenamefont{LaHaye, Buu,
  Camarota, and Schwab}}]{LaHaye:2004p1464}
\bibinfo{author}{\bibfnamefont{M.~D.} \bibnamefont{LaHaye}},
  \bibinfo{author}{\bibfnamefont{O.}~\bibnamefont{Buu}},
  \bibinfo{author}{\bibfnamefont{B.}~\bibnamefont{Camarota}}, \bibnamefont{and}
  \bibinfo{author}{\bibfnamefont{K.~C.} \bibnamefont{Schwab}},
  \bibinfo{journal}{Science} \textbf{\bibinfo{volume}{304}},
  \bibinfo{pages}{74} (\bibinfo{year}{2004}).

\bibitem[{\citenamefont{Regal et~al.}(2008)\citenamefont{Regal, Teufel, and
  Lehnert}}]{Regal:2008p555}
\bibinfo{author}{\bibfnamefont{C.~A.} \bibnamefont{Regal}},
  \bibinfo{author}{\bibfnamefont{J.~D.} \bibnamefont{Teufel}},
  \bibnamefont{and} \bibinfo{author}{\bibfnamefont{K.~W.}
  \bibnamefont{Lehnert}}, \bibinfo{journal}{Nature Phys.}
  \textbf{\bibinfo{volume}{4}}, \bibinfo{pages}{555} (\bibinfo{year}{2008}).

\bibitem[{\citenamefont{Kippenberg and Vahala}(2008)}]{Kippenberg:2008p7}
\bibinfo{author}{\bibfnamefont{T.~J.} \bibnamefont{Kippenberg}}
  \bibnamefont{and} \bibinfo{author}{\bibfnamefont{K.~J.}
  \bibnamefont{Vahala}}, \bibinfo{journal}{Science}
  \textbf{\bibinfo{volume}{321}}, \bibinfo{pages}{1172} (\bibinfo{year}{2008}).

\bibitem[{\citenamefont{Buks and Roukes}(2002)}]{Buks2002}
\bibinfo{author}{\bibfnamefont{E.}~\bibnamefont{Buks}} \bibnamefont{and}
  \bibinfo{author}{\bibfnamefont{M.~L.} \bibnamefont{Roukes}},
  \bibinfo{journal}{J. of Microelectromech. Syst.}
  \textbf{\bibinfo{volume}{11}}, \bibinfo{pages}{802} (\bibinfo{year}{2002}).

\bibitem[{\citenamefont{Karabalin et~al.}(2009)\citenamefont{Karabalin, Cross,
  and Roukes}}]{Karabalin:2009p177}
\bibinfo{author}{\bibfnamefont{R.~B.} \bibnamefont{Karabalin}},
  \bibinfo{author}{\bibfnamefont{M.~C.} \bibnamefont{Cross}}, \bibnamefont{and}
  \bibinfo{author}{\bibfnamefont{M.~L.} \bibnamefont{Roukes}},
  \bibinfo{journal}{Phys. Rev. B} \textbf{\bibinfo{volume}{79}},
  \bibinfo{pages}{165309} (\bibinfo{year}{2009}).

\bibitem[{\citenamefont{Santamore et~al.}(2004)\citenamefont{Santamore,
  Doherty, and Cross}}]{Santamore:2004p398}
\bibinfo{author}{\bibfnamefont{D.~H.} \bibnamefont{Santamore}},
  \bibinfo{author}{\bibfnamefont{A.~C.} \bibnamefont{Doherty}},
  \bibnamefont{and} \bibinfo{author}{\bibfnamefont{M.~C.} \bibnamefont{Cross}},
  \bibinfo{journal}{Phys. Rev. B} \textbf{\bibinfo{volume}{70}},
  \bibinfo{pages}{144301} (\bibinfo{year}{2004}).

\bibitem[{\citenamefont{Lifshitz and Cross}(2003)}]{Lifshitz2003}
\bibinfo{author}{\bibfnamefont{R.}~\bibnamefont{Lifshitz}} \bibnamefont{and}
  \bibinfo{author}{\bibfnamefont{M.~C.} \bibnamefont{Cross}},
  \bibinfo{journal}{Phys. Rev. B} \textbf{\bibinfo{volume}{67}},
  \bibinfo{pages}{134302} (\bibinfo{year}{2003}).

\bibitem[{\citenamefont{Yurke et~al.}(1995)\citenamefont{Yurke, Greywall,
  Pargellis, and Busch}}]{Yurke:1995p872}
\bibinfo{author}{\bibfnamefont{B.}~\bibnamefont{Yurke}},
  \bibinfo{author}{\bibfnamefont{D.~S.} \bibnamefont{Greywall}},
  \bibinfo{author}{\bibfnamefont{A.~N.} \bibnamefont{Pargellis}},
  \bibnamefont{and} \bibinfo{author}{\bibfnamefont{P.~A.} \bibnamefont{Busch}},
  \bibinfo{journal}{Phys. Rev. A} \textbf{\bibinfo{volume}{51}},
  \bibinfo{pages}{4211} (\bibinfo{year}{1995}).

\bibitem[{\citenamefont{Sapmaz et~al.}(2003)\citenamefont{Sapmaz, Blanter,
  Gurevich, and van~der Zant}}]{Sapmaz:2003p389}
\bibinfo{author}{\bibfnamefont{S.}~\bibnamefont{Sapmaz}},
  \bibinfo{author}{\bibfnamefont{Y.~M.} \bibnamefont{Blanter}},
  \bibinfo{author}{\bibfnamefont{L.}~\bibnamefont{Gurevich}}, \bibnamefont{and}
  \bibinfo{author}{\bibfnamefont{H.~S.~J.} \bibnamefont{van~der Zant}},
  \bibinfo{journal}{Phys. Rev. B} \textbf{\bibinfo{volume}{67}},
  \bibinfo{pages}{235414} (\bibinfo{year}{2003}).

\bibitem[{\citenamefont{Nayfeh et~al.}(1995)\citenamefont{Nayfeh, Kreider, and
  Anderson}}]{Nayfeh:1995p387}
\bibinfo{author}{\bibfnamefont{A.~H.} \bibnamefont{Nayfeh}},
  \bibinfo{author}{\bibfnamefont{W.}~\bibnamefont{Kreider}}, \bibnamefont{and}
  \bibinfo{author}{\bibfnamefont{T.~J.} \bibnamefont{Anderson}},
  \bibinfo{journal}{AIAA} \textbf{\bibinfo{volume}{33}}, \bibinfo{pages}{1122 }
  (\bibinfo{year}{1995}).

\bibitem[{\citenamefont{Poot et~al.}(2007)\citenamefont{Poot, Witkamp, Otte,
  and van~der Zant}}]{Poot:2007p392}
\bibinfo{author}{\bibfnamefont{M.}~\bibnamefont{Poot}},
  \bibinfo{author}{\bibfnamefont{B.}~\bibnamefont{Witkamp}},
  \bibinfo{author}{\bibfnamefont{M.~A.} \bibnamefont{Otte}}, \bibnamefont{and}
  \bibinfo{author}{\bibfnamefont{H.~S.~J.} \bibnamefont{van~der Zant}},
  \bibinfo{journal}{Phys. Stat. Sol. B} \textbf{\bibinfo{volume}{244}},
  \bibinfo{pages}{4252} (\bibinfo{year}{2007}).

\bibitem[{\citenamefont{Cleland}(2003)}]{NCleland:2003p509}
\bibinfo{author}{\bibfnamefont{A.~N.} \bibnamefont{Cleland}},
  \emph{\bibinfo{title}{Foundations of nanomechanics: from solid-state theory
  to device applications.}} (\bibinfo{publisher}{Springer},
  \bibinfo{year}{2003}).

\bibitem[{\citenamefont{Verbridge et~al.}(2008)\citenamefont{Verbridge,
  Craighead, and Parpia}}]{Verbridge:2008p013112}
\bibinfo{author}{\bibfnamefont{S.~S.} \bibnamefont{Verbridge}},
  \bibinfo{author}{\bibfnamefont{H.~G.} \bibnamefont{Craighead}},
  \bibnamefont{and} \bibinfo{author}{\bibfnamefont{J.~M.}
  \bibnamefont{Parpia}}, \bibinfo{journal}{Appl. Phys. Lett.}
  \textbf{\bibinfo{volume}{92}}, \bibinfo{pages}{013112}
  (\bibinfo{year}{2008}).

\bibitem[{\citenamefont{Sazonova et~al.}(2004)\citenamefont{Sazonova, Yaish,
  \"Ust\"unel, Roundy, Arias, and McEuen}}]{Sazonova:2004p4487}
\bibinfo{author}{\bibfnamefont{V.}~\bibnamefont{Sazonova}},
  \bibinfo{author}{\bibfnamefont{Y.}~\bibnamefont{Yaish}},
  \bibinfo{author}{\bibfnamefont{H.}~\bibnamefont{\"Ust\"unel}},
  \bibinfo{author}{\bibfnamefont{D.}~\bibnamefont{Roundy}},
  \bibinfo{author}{\bibfnamefont{T.~A.} \bibnamefont{Arias}}, \bibnamefont{and}
  \bibinfo{author}{\bibfnamefont{P.~L.} \bibnamefont{McEuen}},
  \bibinfo{journal}{Nature} \textbf{\bibinfo{volume}{431}},
  \bibinfo{pages}{284} (\bibinfo{year}{2004}).


\bibitem[{foo()}]{footnote}
\bibinfo{note}{The measured responses show a small deviation from the simulated responses due to crosstalk in the measurement equipment.}

\bibitem[{\citenamefont{Postma et~al.}(2005)\citenamefont{Postma, Kozinsky,
  Husain, and Roukes}}]{Postma:2005p1338}
\bibinfo{author}{\bibfnamefont{H.~W.~C.} \bibnamefont{Postma}},
  \bibinfo{author}{\bibfnamefont{I.}~\bibnamefont{Kozinsky}},
  \bibinfo{author}{\bibfnamefont{A.}~\bibnamefont{Husain}}, \bibnamefont{and}
  \bibinfo{author}{\bibfnamefont{M.~L.} \bibnamefont{Roukes}},
  \bibinfo{journal}{Appl. Phys. Lett.} \textbf{\bibinfo{volume}{86}},
  \bibinfo{pages}{223105} (\bibinfo{year}{2005}).

\bibitem[{\citenamefont{Kraus et~al.}(2001)\citenamefont{Kraus, Erbe, Blick,
  Corso, and Richter}}]{Kraus:2001}
\bibinfo{author}{\bibfnamefont{A.}~\bibnamefont{Kraus}},
  \bibinfo{author}{\bibfnamefont{A.}~\bibnamefont{Erbe}},
  \bibinfo{author}{\bibfnamefont{R.}~\bibnamefont{Blick}},
  \bibinfo{author}{\bibfnamefont{G.}~\bibnamefont{Corso}}, \bibnamefont{and}
  \bibinfo{author}{\bibfnamefont{K.}~\bibnamefont{Richter}},
  \bibinfo{journal}{Appl. Phys. Lett.} \textbf{\bibinfo{volume}{79}},
  \bibinfo{pages}{3521} (\bibinfo{year}{2001}).

\bibitem[{\citenamefont{H\"uttel et~al.}(2009)\citenamefont{H\"uttel, Steele,
  Witkamp, Poot, Kouwenhoven, and van~der Zant}}]{huttel:2009}
\bibinfo{author}{\bibfnamefont{A.~K.} \bibnamefont{H\"uttel}},
  \bibinfo{author}{\bibfnamefont{G.~A.} \bibnamefont{Steele}},
  \bibinfo{author}{\bibfnamefont{B.}~\bibnamefont{Witkamp}},
  \bibinfo{author}{\bibfnamefont{M.}~\bibnamefont{Poot}},
  \bibinfo{author}{\bibfnamefont{L.~P.} \bibnamefont{Kouwenhoven}},
  \bibnamefont{and} \bibinfo{author}{\bibfnamefont{H.~S.~J.}
  \bibnamefont{van~der Zant}}, \bibinfo{journal}{Nano Lett.}
  \textbf{\bibinfo{volume}{9}}, \bibinfo{pages}{2547} (\bibinfo{year}{2009}).

\end{thebibliography}
\end{document}